\documentclass[twocolumn, linenumbers]{aa} 

\usepackage{graphicx}
\usepackage{float}
\usepackage{multirow}
\usepackage{subcaption} 
\usepackage{lscape}
\usepackage{txfonts}
\usepackage{xcolor}
\usepackage{placeins}
\usepackage[hidelinks]{hyperref}

\begin{document}

\title{FIP Bias Evolution in an Emerging Active Region as observed in SPICE Synoptic Observations}

\author{
T. Varesano \inst{1,2} 
\and D. M. Hassler\inst{1} 
\and N. Zambrana Prado\inst{3} 
\and J. M. Laming\inst{4} 
\and J. Plowman\inst{1} 
\and D. J. Knipp \inst{2}
\and M. Molnar\inst{1} 
\and K. Barczynski\inst{5,6} 
\and The SPICE consortium
}

\institute{
Southwest Research Institute, Boulder, CO 80302, USA 
\and
Smead Aerospace Engineering Sciences Department, University of Colorado Boulder, Boulder, CO, USA 
\and
University College London, Mullard Space Science Laboratory, Holmbury St. Mary, Dorking, Surrey, RH5 6NT, UK 
\and
Space Science Division, Naval Research Laboratory, Code 7684, Washington, DC 20375, USA 
\and
ETH-Zurich, Hönggerberg Campus, HIT Building, Zürich, Switzerland 
\and
PMOD/WRC, Dorfstrasse 33, 7260 Davos Dorf, Switzerland
}

\date{Received February 2025; accepted December 2025}

\abstract
{}
{We investigate the time evolution of relative elemental abundances in the context of the first ionization potential effect focusing on an active region. Our aim is to characterize this evolution in different types of solar active region structures as well as in different atmospheric layers. We wish to assess how the measured changes relate to different magnetic topologies by computing abundance enhancement in different conditions using the ponderomotive force model.  
} 
{Leveraging SPICE (Spectral Imaging of the Coronal Environment) spectroscopic observations of extreme ultraviolet (EUV) lines from ions formed across a broad temperature range—from the upper chromosphere to the low corona—, we perform relative abundance ratios following differential emission measure analysis. This methodology yields abundance maps from low, intermediate, and high first ionization potential elements.}
{We obtain the temporal evolution of a number of abundance ratios for different structures on the Sun. We compare these results with the outcomes of the ponderomotive force model. We find good correlation between the model and our results, suggesting an Alfvén-wave driven fractionation of the plasma. Fan loops, loop footpoints and AR boundaries exhibit coronal abundances, while the AR core shows more photospheric-like composition. A slow and steady increase in the Mg/Ne FIP bias values is observed, starting around 1.5 and increasing by about 50\% after two days. The S/O evolution coupled with the model brings evidence of resonant waves fractionating the plasma in transition region structures.} 
{}
\maketitle

\section{Introduction} \label{sec:intro}

The growing number of Solar science missions in the last two decades has enabled solar physicists to address a great number of unresolved questions regarding plasma dynamics and mechanisms in the solar atmosphere. Solar Orbiter (SolO), the joint ESA and NASA solar and heliospheric mission, was launched in February 2020 \citep{Muller2020}.

Using the onboard SPICE (Spectral Imaging of the Coronal Environment) instrument, we focus on the first ionization potential (FIP) 
as one of the discriminating factors when it comes to elemental abundance variations. One way to characterize these compositional variations is through the “FIP effect” \citep{meyer_1985, MEYER1991269}, whereby elements with a low first ionization potential (FIP) are enhanced in abundance within certain coronal structures relative to high-FIP elements.

This "FIP effect" is characterized as the ratio between an element's coronal and photospheric abundances (for element $X$, noted $Ab_X^{\text{coronal}}$ and $Ab_X^{\text{photospheric}}$):

\begin{equation}
    \centering
    FIP_{bias} = \frac{Ab_X^{\text{coronal}}}{Ab_X^{\text{photospheric}}}
\end{equation}
Because it is challenging to determine the absolute abundance of elements relative to hydrogen in the upper solar atmosphere, we instead rely on relative abundances. We compare the ratios of radiances between spectral lines from different elements to study their composition.
Several studies have investigated the dependency on this ratio, the `FIP bias', on the type of solar region, magnetic flux, and solar cycle,  and increasing interest has been given to its temporal evolution (see Section 1.2). However, few studies have examined how the FIP bias evolves over time specifically within the transition region or across multiple atmospheric layers — a focus that distinguishes the present work \citep{widing_rate_2001,Baker_2015}. \cite{1995Sheeley}, \cite{Widing_1997}, \cite{1997Young}, and \cite{widing_rate_2001} found photospheric plasma composition in newly emerged loops, indicating that flux emergence provides a potential reservoir of low-FIP bias plasma to mix with the high-FIP bias plasma contained within AR coronal loops. Understanding the mechanisms by which this material enters coronal loops of the AR and the timescales over which plasma mixing occurs, therefore the evolution of the FIP bias, is crucial to investigate the link between the underlying magnetic field and the coronal plasma \citep{Baker_2015}.

Earlier investigations of the FIP effect have mainly focused on the corona \citep{baker2018coronal, Mihailescu_2022, Baker_2022, 2023Mihailescu}. In contrast, studies treating the Sun as an unresolved source found little to no evidence of FIP fractionation at transition region temperatures \citep{laming1995}, and spatially resolved measurements at those temperatures revealed FIP-related signatures only in select structures \citep{1995Sheeley, widing_rate_2001}. With Solar Orbiter’s improved spatial resolution and its closer approach to the Sun, exploring FIP fractionation in the transition region now becomes especially compelling.

As mentioned in \cite{Baker_2015} and references therein, the current thinking suggests that emerging closed loops start with a photospheric composition, which then transitions to a coronal one in the following days. However, the time scales and exact mechanisms for this transformation are not fully understood. \cite{widing_rate_2001} used the \textit{Skylab} spectroheliograph to investigate the Mg VI/Ne VI ratio. Their work concluded that the FIP bias increased shortly after AR emergence ($\Delta t < 1$ day) to reach coronal values after a few days ($\Delta t > 2/3$ days), and a correlation between the sunspot growth and increasing in FIP bias has also been established.
It is thought that emerging loops of photospheric composition reconnect with older structures (i.e. preexisting loops, formed before the flux emergence and exhibiting higher FIP bias),  lead to fractionated plasma being brought up to the corona, most likely through interchange reconnection.The time scale and mechanisms of this mixing is thought to be in the order of days \citep{Baker_2015,widing_rate_2001}.

The aim of this project is to investigate the variability of FIP bias during the emergence phase of an active region, leveraging the unique opportunity provided by the SPROUTS (Synoptic PRogram of Out of RemoTe Sensing observations) to continuously monitor the Sun. The focus will be on the FIP bias evolution over time, across different structural features, and through various layers of the solar atmosphere specifically within the temperature range of $\log_{10}\frac{T}{1\, \text{K}}$=4.8 to 6.0. These changes in elemental abundances are then compared with predictions made by the ponderomotive force model.

\subsection{Fractionation mechanisms}
The ponderomotive force (see \cite{2015Laming, 2019Laming}) provides the most widely adopted explanation of the FIP effect and also the inverse FIP effect (an enhancement in the abundance of high FIP elements). It posits that Alfvén waves generated either in the corona or coming up from the photosphere travel to chromospheric heights where they encounter strong density gradients and thus are refracted and reflected. That change in direction leads to the "ponderomotive force". This force separates the ions from the neutrals -- the ions traveling towards areas of high wave electric energy density, leading to the observed FIP effect. The newly fractionated plasma is then brought upwards by thermal and diffusion processes. For closed loops, the best match with observations comes with waves that are resonant. Alfvén waves propagate along magnetic field lines and can undergo resonance in the chromosphere, where their frequencies match natural oscillation modes of magnetic flux tubes. At resonance, the wave energy is efficiently converted into localized ponderomotive forces acting on the plasma. The wave travel time from one loop footpoint to the other is an integral number of wave half periods, which suggests a coronal origin for the waves. For open fields with no resonance, coronal or photospheric origins cannot be distinguished.

\cite{Réville2021} implement such ideas with substantially more realistic waves/turbulence in open and closed field regions and find good agreement with observations. Using a shell turbulence model, they demonstrate that starting from coronal heating considerations, the wave dynamics lead to FIP fractionation due to the ponderomotive force.

The footpoints and core of active regions are usually the areas that present the strongest FIP bias in Si/S or Fe/S ratios \citep{Baker_2013, 2019Zambrana, Mihailescu_2022}. However, some unusual low fractionation signatures have been observed at  polarity inversion lines, where flux rope formation is thought to occur \citep{Brooks_2022,Mihailescu_2022}. FIP bias values closer to 1 have also been observed in cases of failed eruptions \citep{Baker_2015}.

Previously observed coronal FIP fractionation is believed to be the result of waves generated within the corona \citep{Mihailescu_2022}. The behavior of sulfur - which remains unfractionated - is consistent with the resonant wave activity confined to coronal loops, supporting a coronal origin. Whether a similar mechanism underlies FIP fractionation in the transition region remains an open question. Investigating this could provide valuable information on the origins, evolution, and heating of these structures. Sulfur therefore plays a central diagnostic role in this context.

\subsection{Previous findings on FIP time dependency}
\cite{widing_rate_2001} found that the Mg/Ne FIP bias increases in a linear fashion with the age of the AR. However, they measured the FIP bias solely using intensity ratios, making the results highly sensitive to temperature and density variations, and thus less reliable -- leading to outstandingly high FIP bias values.
The observed evolution of fractionation appears to occur on a timescale consistent with diffusion (on the order of a few days); however, the fractionation process within the chromosphere itself likely proceeds on much shorter timescales. The diffusion timescale primarily reflects the delay in transmitting the chromospheric fractionation signature to the overlying coronal structure. It is also plausible that this behavior varies across different solar structures, particularly if processes such as evaporation or other mass flows are involved.
In agreement with \cite{1995Sheeley}, \cite{1997Young} and \cite{Widing_1997}, emerging loops exhibited photospheric composition, and coronal values of abundances were reached after two or three days. \cite{widing_rate_2001} found abundance enhancements up to a factor of 7 after three days. In the ponderomotive force model for fractionation, \cite{2015Laming} suggests a timescale of about 10$^3$ s for the local abundance anomaly, and several days for observing a global change of composition including the necessary processes of diffusion and thermal transport.

More recently, \cite{Baker_2015} studied a mature AR in its early decay phase, and found that the FIP bias peaks when the active region reaches middle and late decay phases, based on precise monitoring for several sub-regions. In the same fashion, \cite{ko2016correlation} found decreasing FIP bias in decaying ARs. \cite{Baker_2015} and \cite{ko2016correlation} emphasize that magnetic field evolution, especially small-scale flux emergence, is an important actor in compositional changes during AR evolution, as well as density and temperature, which affect abundances at an earlier stage. The high FIP bias is found to be conserved or amplified only in localized area of high magnetic flux density (i.e., in the AR core). In all other areas of the AR, the FIP bias decreased. In younger, smaller active regions, \cite{Baker_2013} found FIP bias values around 2-3. These weaker values might be correlated with the young age of the AR, as well as an eventual reconnection happening with an adjacent coronal hole.

\cite{baker2018coronal} studied emerging flux regions (EFRs), which can be thought of as the coronal counterparts of photospheric magnetic dipoles, and five of the seven regions studied presented a fractionated (FIP bias $>$ 1.5) in less than a day after emergence. They followed up with data from \textit{Skylab}, which showed an increase in the FIP bias at a rate of [0.9 - 2.4] per day for 5 to 7 days for 4 different ARs in their emergence phase.

\cite{to2021evolution} studied long-lived, stable structures and found that over one day, the Ca/Ar ratio depicted a drastic increase while the Si/S ratio stayed constant. It is also worth noticing that cases of inverse-FIP effect \citep{2021LamingInverse} have been observed on the Sun in isolated flare regions \citep{doschek2015anomalous}, where Ar emission was strongly enhanced relative to Ca.

More recently, \cite{Testa_2023} combined Hinode/EIS \citep{2007Hinode} and IRIS \citep{2014iris} observations to study the evolution of the FIP bias within a quiet, slowly decaying active region over 10 days. They observed a modest and stable FIP bias of $\sim$1.5–2 in the AR core. In contrast, persistently high FIP bias values ($\sim$2.5–3.5) were detected in the outflow regions at the AR boundaries and in the fan loops. They found only limited temporal variability on average but more dramatic changes over a timescale of hours.

\subsection{Behavior of intermediate-FIP elements}
Intermediate-FIP ($\approx$ 10 eV) elements, particularly C and S, are important for SPICE abundance determination, and their FIP effect behavior varies with coronal context. 

Their abundances are highly dependent on the solar atmosphere's wave energy and magnetic topology.
For example, S has been observed as behaving as a high-FIP element in closed-field areas \citep{2002Lanzafame, 2020Kuroda} and coronal holes with fractionation values from 1.1 to 1.6 \citep{2019Laming}, whereas in open magnetic field areas, and associated with strong enough magnetic field, S shows a low-FIP element behavior \citep{Giammanco_2007, 2012Schmelz, Reames_2018, 2019Laming,2020Kuroda}. 
Carbon's abundance in coronal holes and closed loops exhibits only a mild enhancement, like S. In open field regions with lower energy fluxes, C behaves more like a low-FIP element with significant fractionation values (2 and above). 
\cite{2019Laming} postulate that in open magnetic field the lack of a resonance means that wave energy can leak out of the corona and develop significant ponderomotive acceleration in the lower chromospheric regions, where the background gas is neutral. In such conditions, S/O and C/O can fractionate. The effect is enhanced with torsional waves due to their lower associated slow mode wave amplitudes. At higher altitudes where the background chromospheric gas is ionized, only the ``true'' low FIP elements are sufficiently highly ionized to fractionate significantly.\\ 

SPICE is the spectrometer with the best coverage of the transition region, where most of the previous instruments focused of higher temperature, coronal emission lines. With those synoptic observations, we hope to make progress in understanding the relationship between plasma composition and the coronal heating process.

We seek to bring elements of an answer to the following questions: 1) Does the FIP bias show a significant evolution during the emergence of an AR? 2) Is the fractionation process driven by waves? 3) Is there evidence that plasma fractionation happens below the corona?

In Section \ref{sec:data}, an extensive description of the dataset is provided. Section \ref{sec:methods} presents the methods used for the diagnostics, and we present the results in section \ref{sec:results}. The comparison with the ponderomotive force model is conducted in section \ref{sec:modeling}. We discuss the implications of the results and potential issues in section \ref{sec:discussion}. Finally, section \ref{sec:conclusion} concludes.

\section{Observations} \label{sec:data}

The SPROUTS observations are a set of synoptic observations outside of the normal Solar Orbiter remote sensing windows, not associated with Solar Orbiter Observing Plans (SOOPs). The observations presented are from December 20th to 22nd 2022 during SolO's fourth remote-sensing window (RSW4). During this time period, Solar Orbiter was at 0.92 AU and had an angular separation with Earth with respect to the Sun of 19 degrees. 

Two rasters were recorded per day. The observations were taken with the 4\arcsec\;slit and an exposure time of 60 seconds, for a total raster recording time of a little more than three hours. The data is available on the SPICE data release webpage\footnote{\url{https://doi.org/10.48326/idoc.medoc.spice.5.0}}. The newest data re-processing includes re-calibration, addressing the radiance value issues raised in \cite{2024Varesano}, as well as burn-in and dark subtraction correction.

\begin{figure}
    \centering
    \includegraphics[width=0.9\linewidth]{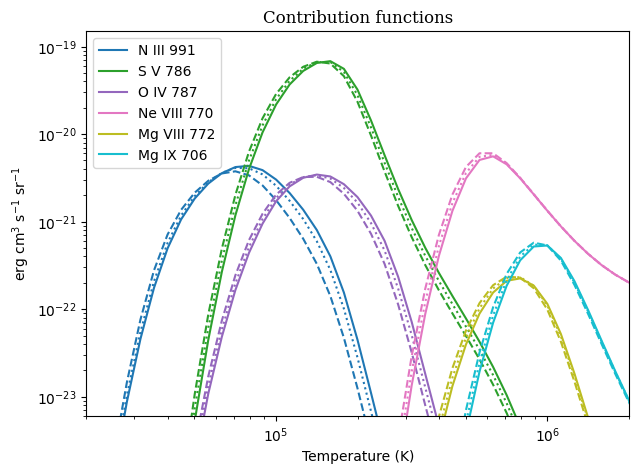}
    \caption{Contribution functions for the SPICE lines of interest computed with the CHIANTI database version 11 \citep{Dufresne2024}. The solid lines are computed with a density of $n_e = 1 \times 10^8$ cm$^{-3}$, dotted lines of $n_e = 1 \times 10^9$ cm$^{-3}$ and dashed lines of $n_e = 1 \times 10^{10}$ cm$^{-3}$.}
    \label{fig:contrib_func}
\end{figure}

Each raster contains 8 windows: O III 703 / Mg IX 706, O II 718, S IV 750 / Mg IX, Ne VIII 770, S V 786 / O IV 787, Ly-gamma-CIII, N III 991 and O VI 1032. Those windows and their average spectra for each raster are plotted in Figure \ref{fig:avspecDec2022}, and the contribution functions for some lines of interest are plotted in Figure \ref{fig:contrib_func}. These contribution functions were computed at three different densities using version 11 \citep{Dufresne2024} of the atomic database CHIANTI and the corresponding IDL routines in SolarSoftware.

Two Active Regions (ARs) were captured in the dataset from December 2022: NOAA 13171 and NOAA 13169. Both of them are registered as $\beta$-class ARs. Their characteristics are summarized in Table \ref{CharacAR}. 

\begin{table}[htbp]
\begin{center}
    \begin{tabular}{lll}
    \hline \hline
         & NOAA 13169 & NOAA 13171\\ \hline
        Hale class & $\beta$ / $\beta$-$\gamma$ & $\beta$ \\
        Start date & 2022/12/17 & 2022/12/19\\
        End date & 2022/12/30& 2022/12/31\\
        Max. sunspot area & 490 & 250 (millionths)\\
        Max. nbr of spots & 30 &  12 \\
        Associated flares  & 3 M-class & 6 C-class \\
        \hline \hline
    \end{tabular}
    \caption{Technical description of observed ARs.} \label{CharacAR}
\end{center}
    \label{tab:class_AR}
\end{table}

\begin{figure}
    \hspace*{-0.5cm}
    \centering
    \includegraphics[width=\linewidth]{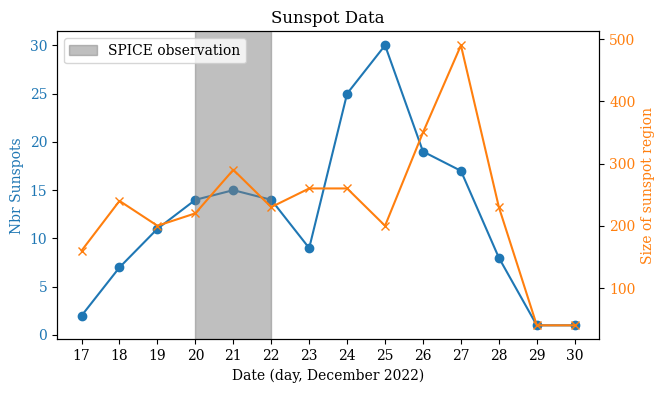}
    \caption{Evolution of AR 13169 in terms of sunspot number (blue curve) and sunspot area (orange curve). Data is the courtesy of \url{https://www.spaceweatherlive.com}. During this period, 3 M-class flares and 61 C-class flares have been recorded.
    The time of the observations studied here is highlighted in gray.}
    \label{fig:sunspot_evo}
\end{figure}

\begin{figure}
    \hspace*{-0.5cm}
    \centering
    \includegraphics[width=1\linewidth]{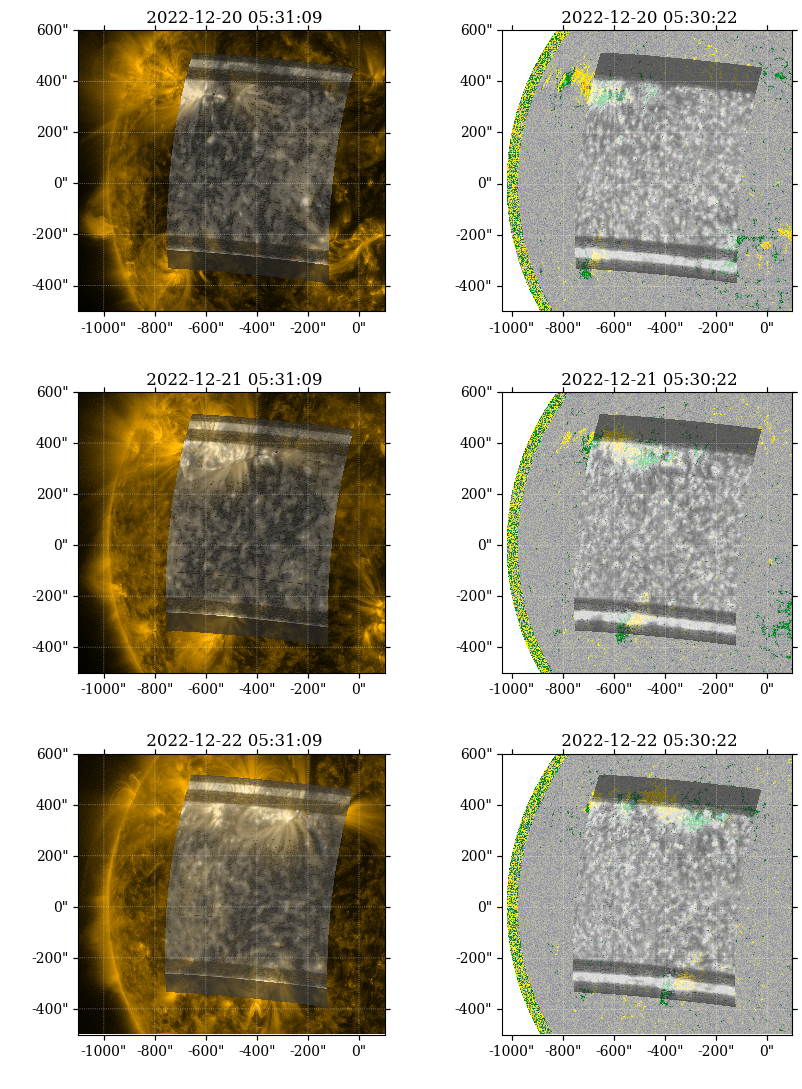}
    \caption{Evolution of AR 13169 (situated in the top left on the first raster) from 2022 December 20 to 2022 December 22. Left: SPICE's Ne VIII 770 \AA\;rasters overlayed on SDO/AIA 171 \AA. Right: corresponding SDO/HMI magnetograms with C III 977  \AA\;rasters. On the magnetograms, green/yellow areas represent positive/negative polarity. 
        Note: the dark horizontal lines ("dumbbells") are square alignment apertures, located at each end of the slits
    .}
    \label{fig:aiahmi}
\end{figure}

NOAA 13169 stays the longest in SPICE's FOV, and therefore has been targeted for the rest of the analysis. The AR was first observed on December 17, 2022, already presenting a bipolarity in its sunspot structure. During its observation by SPICE, the AR is within its late, slow emerging phase (see Figure \ref{fig:sunspot_evo}). Its magnetic configuration is complex enough that opposite polarity spots cannot be clearly distinguished. Figure \ref{fig:aiahmi} shows the evolution of the AR seen in the SDO/AIA \citep{Lemen2012} 171 \AA\; band, and the corresponding magnetograms from HMI \citep{2012Scherrer}. The latter are plotted within the range of $\pm$ 200 G. The leading positive polarity sunspot group stays consistent through the observations, bordered by a negative, more dispersed field on its west side.

Two main outflows are seen in the EUV FOV. On December 22, a full loop can be observed within SPICE's raster (see Figure \ref{fig:aiahmi}, bottom panel). This loop seems to be newly formed, linking the very strong positive magnetic field area to the negative area. Further confirmation of high magnetic activity is provided by the flares associated with the AR (on December 20, one M1.1-class flare at 13:59 UT and eight C-class flares were associated with AR13169).

\begin{table*}[htbp]
\centering
\begin{tabular}{llllll}
\hline \hline
Spectral lines   & logT (K) & FIP (eV) & Transition   & Sub-windows ($\AA$)  \\ \hline
\ion{O}{i} 988.6              & 4.2      & 13.6     & 2s$^2$ 2p$^4$ $^3$P$_2$ - 2s$^2$ 2p$^3$ $^2$D$_0$   & 987.4 -- 989.1         \\
\ion{H}{i} 972.5              & 4.5      & 13.6     & 1s$^2$ $^2$S$_{1/2}$ - 5p $^2$P$_{1/2}$   & 969.0 -- 974.5           \\
\ion{O}{ii} 718 (mult.)        & 4.7      & 13.6   & 2s$^2$ 2p$^3$ $^2$D$_{5/2}$ - 2s 2p$^4$ $^2$D$_{5/2}$   & 716.3 -- 720.9    \\
\ion{C}{iii} 977               & 4.8      & 11.3    & 2s$^2$ $^1$S$_0$ - 2s 2p $^1$P$_1$   & 974.5 --  980.3       \\
\ion{N}{iii} 989 *             & 4.8      & 14.5     &  2s$^2$ 2p $^2$P$_{1/2}$ - 2s 2p$^2$ \, $^2$D$_{3/2}$       & 989.1 -- 990.6         \\
\ion{N}{iii} 991 *         & 4.8      & 14.5     & 2s$^2$ 2p $^2$P$_{3/2}$ - 2s 2p$^2$ \, $^2$D$_{3/2}$     & 990.6 -- 993.4       \\
\ion{O}{iii} 702.8 *       & 4.9      & 13.6     & 2s$^2$ 2p$^2$ $^3$P$_{1}$ - 2s 2p$^3$ $^3$P$_{0}$ & 700.3 -- 703.1       \\
\ion{O}{iii} 703.8 *             & 4.9      & 13.6     & 2s$^2$ 2p$^2$ $^3$P$_{2}$ - 2s 2p$^3$ $^3$P$_{1}$ & 703.1 -- 704.8     \\
\ion{S}{iv} 750 *               & 5.0      & 10.4     & 3s$^2$ 3p $^2$P$_{3/2}$ - 3s 3p$^2$ $^2$P$_{3/2}$ & 749.8 -- 752.5        \\
\ion{S}{v} 786 *              & 5.2      & 10.4     & 3s$^2$ $^1$S$_{0}$ - 3s 3p $^1$P$_{1}$            & 784.2 -- 787.0         \\
\ion{O}{iv} 787 *                & 5.2      & 13.6     & 2s$^2$ 2p $^2$P$_{1/2}$ - 2s 2p$^2$ $^2$D$_{3/2}$ & 787.0 -- 789.8       \\
\ion{O}{vi} 1032 *                & 5.4      & 13.6     & 1s$^2$ 2s $^2$S$_{1/2}$ - 1s$^2$ 2p $^2$P$_{3/2}$ & 1029.1  --  1034.9    \\
\ion{Ne}{viii} 770 *          & 5.8      & 21.6     & 1s$^2$ 2s $^2$S$_{1/2}$ - 1s$^2$ 2p $^2$P$_{3/2}$ & 767.2 -- 771.5         \\
\ion{Mg}{viii} 772              & 5.9      & 7.6      & 2s$^2$ 2p \;$^2$P$_{3/2}$ - 2s2p$^2$\;$^4$P$_{5/2}$   & 771.5 -- 775.2    \\
\ion{Mg}{ix} 749               & 6.0      & 7.6      & 2s$^2$ 2p $^1$P$_1$ - 2p$^2$ $^1$D$_{2}$            & 747.8 -- 749.8    \\
\ion{Mg}{ix} 706            & 6.0      & 7.6      & 2s$^2$ $^1$S$_{0}$ - 2s 2p $^3$P$_{1}$            & 704.85 -- 708.1    \\ \hline \hline
\end{tabular}%
  \caption{Lines extracted from the SPROUTS dataset, and the wavelength window used to determine the Gaussian parameters for the fitting. The lines marked with * have been used in the DEM computation.}
\label{tab:compo_study}
\end{table*}

    To investigate the plasma composition and dynamics within the observed active region, we now turn to the spectroscopic diagnostics and data processing methods used in this study.


\section{Methods} \label{sec:methods}

Figure \ref{fig:radmapDec2022} shows C III 977\,\AA, O VI 1032\,\AA\;and Ne VIII 770\,\AA\, radiance maps. All radiance maps were obtained by fitting the spectral lines with a Gaussian model, using sub-windows and multiple Gaussian curves (up to three) if several emission peaks were present in a single window (see Table \ref{tab:compo_study}). 

 Note that the O II 718\,\AA\; and O VI 1032\,\AA\; are the only windows containing a single emission peak, while S IV 750\,\AA\; / Mg IX 749\,\AA\;, Ne VIII 770\,\AA\; and Ly-gamma-CIII containing two distinct emission peaks. The last three windows, O III 703\,\AA\; / Mg IX 706\,\AA\;, S V 786\,\AA\; / O IV 787\,\AA\; and N III 991\,\AA\; present three emission lines.

\begin{table}[htbp]
\hspace*{-0.5cm}
\centering
\begin{tabular}{lll}
\hline \hline
Line               & Radiances QS     &  Radiances AR  \\ \hline
\ion{O}{iii} 702.8 & 18.1 $\pm$ 0.7  & 67.1 $\pm$ 0.4   \\
\ion{O}{iii} 703.8 & 17.8 $\pm$ 0.4  & 78.5 $\pm$ 0.4     \\
\ion{Mg}{ix} 706   & 2.8 $\pm$ 0.5  & 42.4 $\pm$ 0.6     \\
\ion{Mg}{ix} 749   & 3.1 $\pm$ 0.2   & 12.8 $\pm$ 0.2    \\
\ion{S}{iv} 750    & 3.3 $\pm$ 0.2   & 20.5 $\pm$ 0.1     \\
\ion{Ne}{viii} 770 & 33.7 $\pm$ 0.3  & 495.5 $\pm$ 0.2   \\
\ion{Mg}{viii} 772 & 2.7 $\pm$ 0.5   & 43.6 $\pm$ 0.7     \\
\ion{S}{v} 786     & 23.4 $\pm$ 1.4  & 125.3 $\pm$ 0.7     \\
\ion{O}{iv} 787    & 48.8 $\pm$ 0.4  & 224.7 $\pm$ 0.6    \\
\ion{C}{iii} 977   & 590.0 $\pm$ 8.0 & 3922.9 $\pm$ 26.0 \\
\ion{N}{iii} 989.8 & 24.8 $\pm$ 1.2 & 135.3 $\pm$ 1.4     \\
\ion{N}{iii} 991.5 & 36.1 $\pm$ 1.0 & 383.6 $\pm$ 1.8    \\
\ion{O}{vi} 1032   & 189.4 $\pm$ 0.6 & 2318.2 $\pm$ 2.0 \\ \hline \hline
\end{tabular}
\caption{Radiance values (mW/m$^2$/sr) from December 20 2022, 04:34 UT for a Quiet Sun macropixel (QS) of 12600 pixels and an Active Region (AR) macropixel of 1200 pixels.} \label{tab:radiances}
\end{table}

To compute the relative FIP bias (the ratio of FIP bias of element X over element Y rather than H), we used three different line ratios, covering different ranges of temperatures, and therefore regions in the solar atmosphere: S V 786 \AA\;(low-FIP proxy, 10.4 eV, $\log(T)=5.2$ K) and O IV 787 \AA\;(high-FIP, 13.6 eV, $\log(T)=5.2$ K), covering the lower transition region; and Mg VIII 772 \AA\; (low-FIP, 7.6 eV, $\log(T)=5.85$ K) and Ne VIII 770 \AA\; (very-high FIP, 21.6 eV, $\log(T)=5.8$ K) as well as Mg IX 706 \AA\; (low-FIP, 7.6 eV, $\log(T)=5.9$ K) and Ne VIII 770 \AA, covering the lower corona.

Even though sulfur is considered as intermediate-FIP, we still use it in our diagnostics to investigate the behavior of this ambiguous element, knowing that it exhibits different behaviors based on magnetic configuration.

To account for contribution function differences as well as temperature and density effects, we included a differential emission measure (DEM, provides a measure of the amount of the emitting plasma along the line of sight as a function of temperature) factor in our FIP bias computation. 
We do later derive DEMs from SPICE. Also DEM variance are too large for a reference DEM to be acceptable. Real computed DEMs should instead be used. 
The FIP bias is thus numerically computed as
\begin{equation}
    FIP_{bias} = \frac{I_{LF}}{I_{HF}} \left(\frac{Ab^P_{LF}}{Ab^P_{HF}} \frac{<C_{LF} \cdot \text{DEM}>}{<C_{HF} \cdot \text{DEM}>} \right) ^{-1}
    \label{eq:fip}
\end{equation}
Where $LF$ and $HF$ denote respectively the low-FIP and high-FIP elements, and $<a\cdot b>$ designates the scalar product between $a$ and $b$.\\

Determining the DEM from observed radiances in multiple spectroscopic lines is a challenging task due to the limitations of integral inversion methods, such as data insufficiency and the ill-conditioned nature of the problem, particularly in the density dimension\citep{1976CraigBrown, 1997Judge}. Most DEM determination techniques require prior density measurements, and inversion methods often struggle to accurately resolve the general shape and finer details of the DEM, especially for multithermal plasmas, where solutions are biased to specific temperature ranges \citep{2012Testa, 2012Guennou}. However, we use here \cite{Plowman_2020}'s method, and this enables a realistic estimation of the DEM directly from the same data used to compute the radiance, allowing for a consistent determination of the FIP bias as well. We checked that variations in the density chosen (here, $n_e= 10^9$ cm$^{-3}$) do not affect significantly the contribution function. Once the DEM estimates are computed, they are used in Equation \ref{eq:fip}.

\begin{figure*}[!ht]
    \centering
    \includegraphics[width=\linewidth]{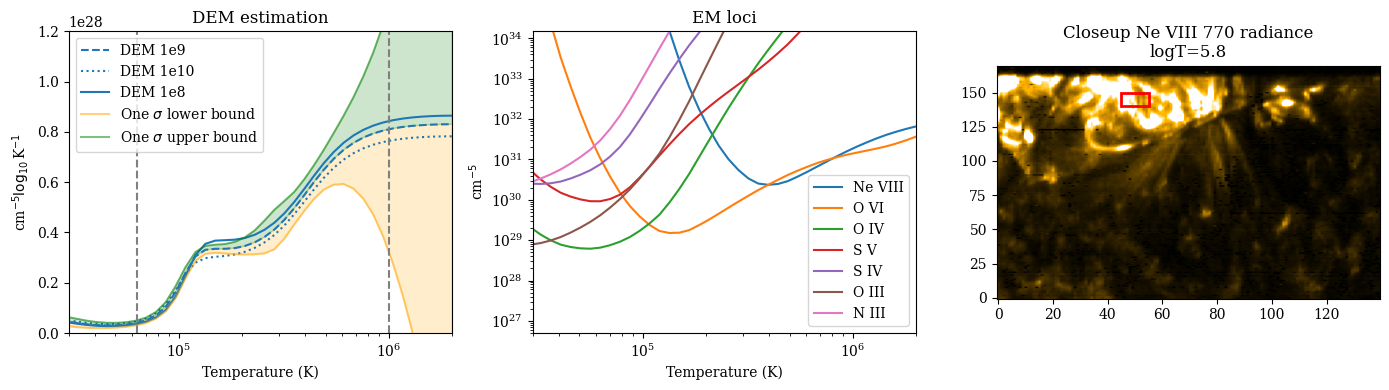}
    \caption{DEM estimation (left panel), EM loci (middle panel, providing an estimation of the upper limit of the DEM) and corresponding zone on the Ne VIII 770 \AA\;map. The DEM estimation indicates a rather multithermal emission. The gray dotted lines indicate the range of temperature where the DEM estimation is reliable, and different line styles represent different densities. Note: the scale for the DEM is linear.}
    \label{fig:dem_em_loci}
\end{figure*}

In Figure \ref{fig:dem_em_loci} are presented the results of the Emission Measure (EM) loci and DEM for the right footpoint of AR 13196 on December 21, 2022 at 03:34 UT. The EM loci shows an intersection at logT = 5.8 K, further confirmed by the DEM showing that most of the contribution comes from material around logT = 5.8 K. The DEM, as well as the contribution functions, do not show a strong density dependency for the lines used. A detailed explanation on how the uncertainties (depicted as the green and orange shaded areas) are computed for this study is presented in Appendix \ref{sec:uncertainties}.

    Now that we have established the methods used, those diagnostics are applied to derive and investigate the spatial and temporal changes in the FIP bias.

\section{Results} \label{sec:results}

Different diagnostics, involving a wide range of temperatures and ionization energies, have been computed. The ratios corresponding to different solar atmospheric layers have been computed using the methods described in Section \ref{sec:methods}.

\begin{figure*}[!ht]
    \centering
    \includegraphics[height=20.8cm]{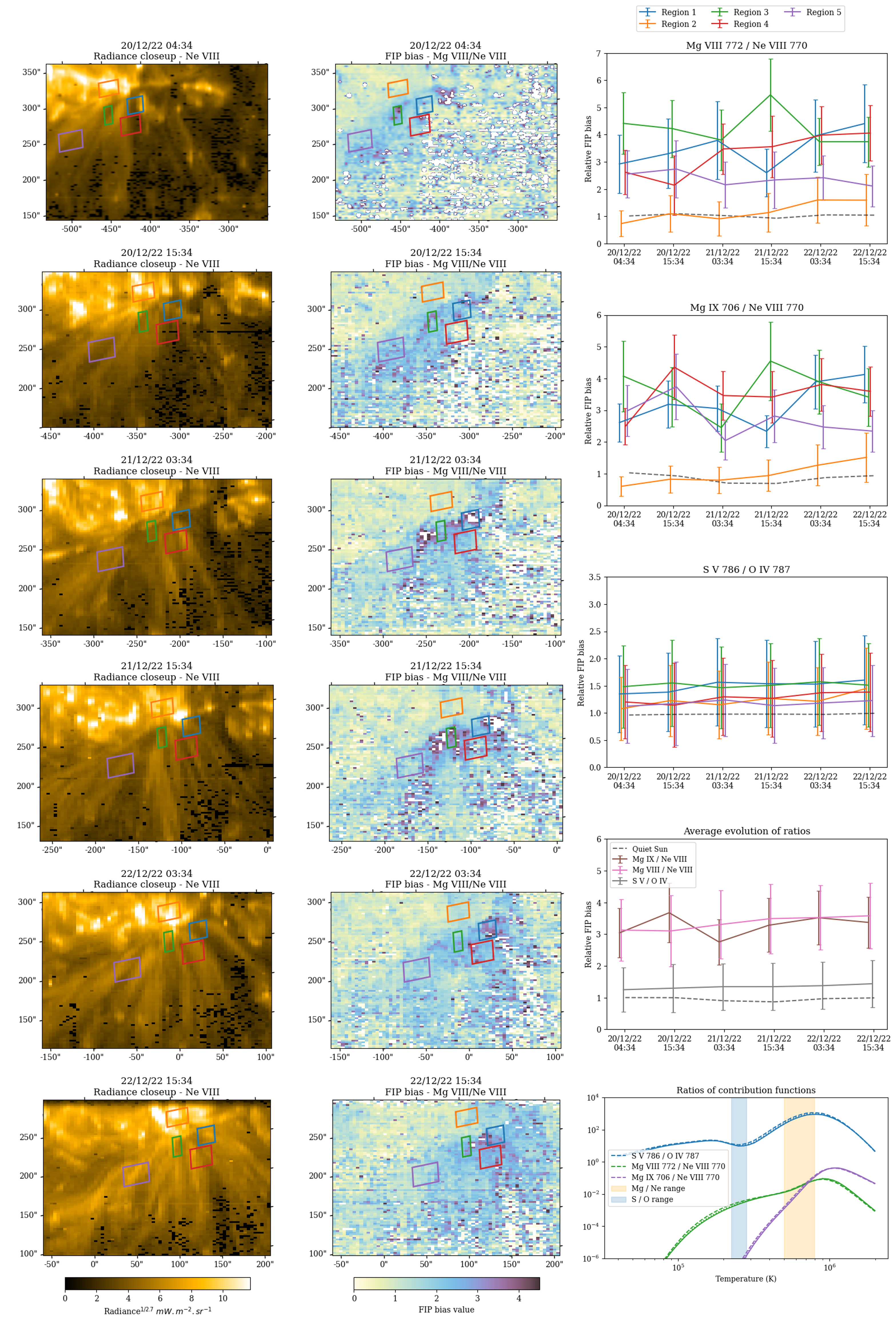}
    \caption{Evolution of AR 13169 with time progressing downwards. Left column: SPICE radiance maps seen in the Ne VIII 770 line ($\log_{10}\frac{T}{1\, \text{K}}$=5.8). Middle column: corresponding FIP bias maps seen in the Mg IX / Ne VIII ratio. Right column: Evolution of the FIP bias values, tracking different zones. Note: time progresses from left to right in this column's plots, 
    and the offset is to show error bars with more clarity. The black dotted line represents the photospheric / quiet Sun reference. Bottom right plot: ratio of contribution functions for the lines presented. The solid lines are computed with a density of $n_e= 10^8$ cm$^{-3}$, dotted with $n_e= 10^9$ cm$^{-3}$ and dashed $n_e= 10^{10}$ cm$^{-3}$.}
    \label{fig:FIP_bias_time}
\end{figure*}

AR 13169 has been tracked by using several macropixels, designated as the color boxes in Figure \ref{fig:FIP_bias_time}. 
Since the rasters are recorded over several hours, the boxes follow the rotation of the Sun rather then specific features of the AR at a given time.

The five distinctly colored boxes (regions) have the following properties for the three-day time series:  1) all stay within the raster bounds with sufficient signal-to-noise ratio to illustrate FIP bias behaviors, and 2) each similarly colored box (region) represents a targeted evolving feature within the AR. Region 1 (blue) is targeted towards the AR footpoint; region 2 (orange) represents the core of the AR; regions 3, 4 and 5 (green, red and purple, respectively) represent fan loops, at a varying distance from the core. In the right column of Fig \ref{fig:FIP_bias_time} the marker's colors correspond to the tracked regions. According to fractionation models, the footpoints are where most of the fractionation would happen; fan loops are more likely subject to interchange reconnection; and the AR core has the most dynamic, unpredictable environment.

While the FIP bias maps are calculated on a pixel-by-pixel basis, we used the average spectrum over the selected region to reduce uncertainties in the right-most panel, which presents a more quantitative FIP bias estimate.

Results of the tracking are shown in Figure \ref{fig:FIP_bias_time}. We begin by discussing the middle panels (i.e., the Mg VIII / Ne VIII maps), followed by the panels in the right column of this figure, and finally an analysis in the context of potential mechanisms at play.

The FIP bias maps seen in the Mg VIII / Ne VIII ratio show well the spatial distribution of the FIP bias within the AR. Whereas the core is mainly photospheric, the footpoints, fan loops and boundary (especially the right-hand one) exhibit strong fractionation. Over the course of the two days, the FIP bias seems to increase and intensify at the AR footpoint, between regions 1 and 3. 
These maps allow us to differentiate between regions where we can observe an enhancement of the different abundance ratios. They inform our selection of different regions to study more closely in a more quantitative fashion. This is illustrated in the plots on the right-hand panel:

Mg VIII / Ne VIII:
region 2 (orange curve) is exhibiting photospheric-like abundances at the beginning of the observations, showing a slow increase over time. The values from the AR core are close to the quiet Sun reference, then shift towards coronal abundances.
The macropixels targeting the AR boundary and fan loops, however, exhibit coronal-like abundances, especially for the ones close to the bright core. They show a global increasing trend, starting from around 2.5 to above 3.5 after two days, except for region 5, which is the furthest from the core.

Mg IX / Ne VIII:
All regions show similar behavior as Mg VIII/Ne VIII. There is however more variation in the values, which we discuss later in this section.

S V / O IV:
this diagnostic is particularly interesting because both lines are close in wavelength and temperature, while exhibiting a substantial FIP gap as well as a good signal-to-noise ratio. All regions show very correlated trends as well as close values, starting around photospheric figures and showing a slight increase over the dates studied.

Overall, the ratios show either a consistent, slow increase in the FIP bias over time or stay constant at coronal values. The slow rate of increasing FIP bias matches the fact that the AR 13169 is in a slow emerging phase, so that the waves are trapped and resonate within the loop. The results suggest that the resonant frequency of the loop does not change dramatically, therefore waves have a chance to remain in resonance. 
The fact that for most ratios the evolution of the FIP bias is structure-dependent could be linked to what \cite{warren2016transition} identified, which is that intense heating events show close to photospheric composition, while more stable structures like coronal "fans" are consistent with the usual coronal composition. 

An important caveat to be noted is, as shown by the bottom right plot in Figure \ref{fig:FIP_bias_time}, the dependence of the inferred FIP bias on the ratio of the contribution functions. For the Mg/Ne and S/O pairs, the ratio of contribution functions varies noticeably with temperature. While it remains relatively constant over the range considered for S/O and Mg VIII/Ne VIII, the case of Mg IX/Ne VIII shows a stronger temperature dependence. This sensitivity partly reflects an instrumental limitation of SPICE, as the relevant lines peak at different temperatures and the DEM becomes poorly constrained above $\log_{10}\frac{T}{1\, \text{K}}$= 5.8 when only high-FIP lines are available. Consequently, some of the apparent variations in the Mg/Ne ratios may be partly thermal in origin rather than purely compositional. The impact of this temperature dependence on the derived FIP bias is mitigated through the use of the DEM in the computation, rather than relying solely on the direct ratio of the contribution functions.\\

The boundaries / fan loops of the AR show on average a much higher fractionation. The most likely explanation, is the interchange reconnection that might happen between open field lines with photospheric abundance reconnecting with coronal-abundance loops, therefore enriching the coronal composition of the boundaries. Another phenomenon that we could be observing is the very dynamic core of the AR, having emerging loops filling quickly with photospheric abundances (chromospheric plasma) and getting fractionated after a period of time. The high values at the fan loops / footpoints could be the signature of the fractionated plasma staying at the top of the loops for a few hours, then cooling and going back down along the loop charged with coronal abundances.

The enhancement of sulfur compared to O denotes a low-FIP behavior from S. This trend is specifically notable at the footpoints of loops. \cite{2020Kuroda} suggested that the enhancement of intermediate-FIP elements was taking place in the lower chromosphere, where H is not ionized. This phenomenon is even more pronounced in the areas where non-resonant waves are present, which is a characteristic of open-field regions.

\section{Modeling} \label{sec:modeling}
Following the model developed in \cite{2017Laming, 2019Laming}, FIP fractionation is estimated for the active region observed. This model describes the observed fractionation of elemental abundances in the solar atmosphere at a low time resolution, i.e. ignoring the potential bursts of fractionation caused by nanoflares for long-lived structures like coronal loops, active regions, and solar wind outflow regions. Within these static scenarios, fractionation in closed loops is thought to take place in the upper chromosphere.

\begin{figure*}[!ht]
    \centering
    \includegraphics[width=\linewidth]{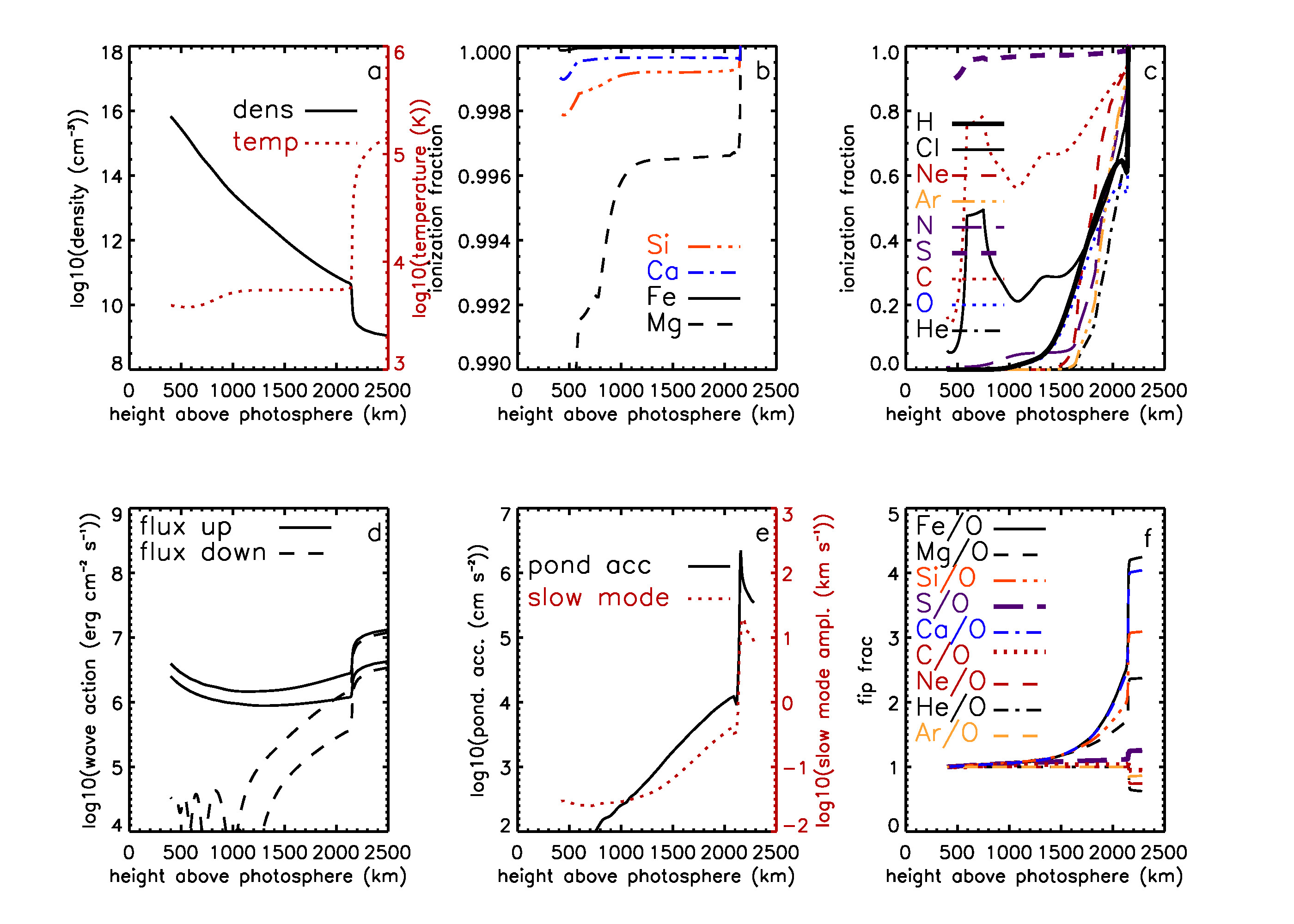}
    \caption{Top panels, from left to right: chromospheric density and temperature structure, low FIP ionization fractions and high FIP ionization fractions. Bottom panels, from left to right: wave energy fluxes, ponderomotive acceleration and slow mode wave amplitudes, and fractionations with height in the chromosphere. While the Fe/O ratio increases, the S/O does not. This behavior is typical of resonant waves on closed loops.}
   
    \label{fig:modelling2}
\end{figure*}

The ponderomotive force acts primarily on the low-FIP elements, which have a significantly higher ionization fraction than high-FIP elements (see top middle and right panels in Figure \ref{fig:modelling2}). This force preferentially lifts them toward the top of the transition region. Upon reaching this region, two key changes occur: the temperature rises enough to fully ionize both low and high FIP elements, eliminating the distinction between ions and neutrals, and the density gradient becomes smaller, causing the effects of the ponderomotive force to vanish. As a result, the fractionation process ceases at the transition region's upper boundary, locking in the established fractionation pattern. Other mechanisms then are thought to transport the fractionated plasma into the corona.

The parameters taken consists of a loop 80,000 km long and with a 100 G magnetic field. Those values are a rough estimate inferred from AIA images. Waves have angular frequencies of 0.332 rad/s and 0.665 rad/s (fundamental and first harmonic), with amplitudes chosen to give FIP fractionations around 4 with respect to oxygen.

The modeled FIP fractionation values are represented in Figure \ref{fig:modelling2}. The key feature is on the right panel, where low-FIP elements start to behave significantly differently with respect to the high-FIP elements in the upper chromosphere. The intermediate element sulfur, which supposedly behaves as a high-FIP element in closed magnetic structures (see \cite{2024Baker} and references therein), stays at a constant ionization rate. However, it is interesting to note that in the ponderomotive force model, S regarding to O qualitatively looks a low-FIP element, being enhanced rather than depleted at a chromospheric altitude of 220 km, even though by not very much (see Figure \ref{fig:modelling2}).

Overall, the data matches the model, similarly to \cite{2023Mihailescu}'s work, where they also compared the ponderomotive force model to observations from Hinode / EIS. Comparing the model and Figure \ref{fig:FIP_bias_time}, the modeled (final) FIP bias for Mg/Ne is 3.20, and S/O show a modest value of 1.26. S/O observations and model show the best agreement, while the Mg/Ne ratios show consistency mostly with regions 1,3 and 4 (AR boundary).

Even though there are still discrepancies between models and observations for changes in abundances, both converge more and more in recent years. These first results of SPICE SPROUTS’ rasters suggest that fractionation happens at a chromospheric level and that the fractionated plasma is brought at the top of the transition region after a time delay of about 24 hours. 
Understanding the differences between FIP bias diagnostics and the mechanisms of the fractionation of S will be crucial for linking in situ plasma observations to their solar surface origin, as S is a frequent observation in both remote sensing and in situ FIP bias diagnostics.


\section{Discussion} \label{sec:discussion}

Investigating the FIP effect and its evolution regarding different variables is essential to understand the fractionation mechanisms of photospheric plasma. From SPICE's observations and our diagnostics, we can conclude that waves are highly likely to cause fractionation. The fact that the Mg/Ne increases is what is expected if the waves causing the fractionation are resonant with the coronal loop, which almost certainly means that they are generated in the coronal part of the loop subject to the boundary conditions at the loop footpoints, analog to an optical resonating cavity. 

Plasma mixing, interchange reconnection, flux emergence and cancellation are important factors in the observed values of FIP bias \citep{Koukras_2025, Testa_2023}. 

We will base our conclusions solely on the Mg VIII / Ne VIII and the S/O ratios. The differences between the two Mg/Ne ratios may stem from a slight temperature mismatch: Mg VIII and Ne VIII are formed at similar temperatures and have nearby wavelengths, whereas Mg IX forms at a higher temperature, near the upper limit of SPICE’s sensitivity. To better constrain the higher-temperature lines, coordinated observations with Hinode/EIS and SPICE would be ideal \citep{Brooks_2022, 2024Brooks}.

During the emergence phase (early in the study), the spectra could be affected by absorption by intervening H I, while those later on are not, presumably because enough of the loop has emerged to sufficiently higher altitudes. This would mainly affect the lines C III 977 and N III 991, which are shortward of the H I edge at 912 \AA\ and can be absorbed \citep{1976Orrall}. Mg VIII/Ne VIII and S V/O IV are likely not affected by this phenomenon, because the lines are so close in wavelength that the H I absorption would affect them both the same, having no impact on the FIP bias computation. The Mg IX 706/Ne VIII 770 have H I absorption cross sections of 3.2 and 4.0 MB respectively, so selective absorption of Ne VIII 770 over Mg IX 706 in a column density of neutral hydrogen of $1.25\times 10^{18}$ cm$^{-2}$ could give rise to an ``apparent FIP effect''. There is no evidence for such an H column in any of the images, and we consider that uncertainties in the DEM reconstruction (see Figure 4, left panel) are a more likely cause of uncertainty.

Observing different behaviors for different magnesium lines could be explained by the difference in temperature and therefore contribution functions, which are a crucial part of estimating DEMs and computing the FIP bias. Further observations and analysis would be useful to assess which ratio is the most reliable, knowing that Mg IX usually has a better signal-to-noise ratio, but Mg VIII has a closer temperature and wavelength to Ne VIII. 

In their study, \cite{widing_rate_2001} postulated that coronal loops emerge with photospheric abundances, and only become coronal-abundant when at least a part of the loop reaches the corona. With this postulate and recent results, including the observations made in this work, the FIP effect observed in this scenario is likely caused by a cascade of events starting from the corona.

The slow temporal evolution over timescales of tens-of-hours to days suggests steady heating, but more chaotic evolution within smaller timescales is likely to happen. Outflow regions maintain consistently high FIP bias throughout the observation period as observed in \cite{Testa_2023}. They found a moderate correlation (Pearson’s correlation of 0.35 for their dataset) between the chromospheric turbulence inferred from IRIS and high FIP bias values at the boundary of the AR, which is what we observe with the lower temperature lines from SPICE. However, our diagnostics yield predominantly photospheric abundance values in the AR core, in contrast to the coronal-like values reported by \cite{Testa_2023}, even though both studies reveal a similarly stable temporal trend.



In Figure \ref{fig:FIP_bias_time} the AR 13169 is emerging slowly. We infer the waves in the magnetic field are trapped in the structure and resonate as in an optics resonant cavity, giving the FIP bias time to develop and increase. Looking at the evolution of the magnetic structure over the set of dates studied, we observe numerous flux cancellation events around the right footpoint of AR 13169. These phenomena could lead to the emergence of new photospheric flux, leading to the lower values observed in Figure \ref{fig:FIP_bias_time}. Combining the SPICE observations with models, it appears that the FIP bias is dependent also on height in the chromosphere and lower transition region. However, once the plasma reaches the transition region, most of the fractionation has happened and the remaining evolution will be very slow. The S V/O IV ratio indicates the presence of resonant waves on the transition region temperature structures, and those waves must be generated in the corona.

\textit{Limitations of the method} 
The interpretation of the SPICE- derived FIP bias values involves several sources of uncertainty. There is uncertainty in the data itself, which has to be propagated through the line fitting and in the DEM estimation. The contribution functions also have their own uncertainty. While the data suggest a slow temporal evolution of the FIP bias, this trend cannot be confirmed with certainty given the limited cadence and possible height-dependent effects in the chromosphere and transition region. A detailed explanation of how we deal with those uncertainties and how we quantify them can be found in Appendix \ref{sec:uncertainties}.
The temperature and shape mismatch in the contribution functions may also contribute to differences in the inferred FIP bias, especially for Mg IX which forms at higher temperatures near the upper sensitivity limit of SPICE. During the emergence phase (early in the study) spectra may have been affected by H I absorption for lines shortward of 912 \AA. However, no clear evidence of large neutral hydrogen columns is found; moreover, once the loops are established, this phenomenon should be minimized. 

It is beyond the scope of this paper to quantitatively relate the rate of temperature increase to the product of wave energy and damping rate. Indeed, there is currently no consensus on the dominant wave damping mechanisms or their efficiency. 
Coronal loops can be heated by episodic events such as nanoflares \citep{2016Dahlburg}. These can in turn excite Alfvén waves. This modulation of the Alfvén wave spectrum would thus influence the ponderomotive force responsible for the FIP fractionation. 
While individual nanoflares occur on timescales of minutes, the resulting changes in the wave environment and ion–neutral coupling in the chromosphere could act cumulatively, producing observable compositional evolution on much longer timescales (tens of hours to days).

Finally, the derived FIP bias values depend on assumptions about plasma homogeneity, ionization equilibrium, and accurate line formation modeling, with uncertainties propagating from the data reduction through DEM estimation. Despite these limitations, the observed trends remain meaningful and provide valuable insight when considered alongside models and complementary observations.

\section{Conclusion} \label{sec:conclusion}

These SPICE SPROUTS observations provide needed insight into the evolution of the FIP bias, specifically looking deeper in the atmosphere to investigate the chromospheric origin of plasma fractionation. The large number of spectral lines recorded and the recurrent recordings enable us to follow the fractionation of elements from the chromosphere to the corona and in time.

We have investigated the variability of FIP bias during the emergence phase of AR 13169. We found clear evidence of resonant waves generated in the corona at the boundaries and footpoints of the AR, and those in the transition region lines. From this study, we conclude that resonant waves are the principal driver of the FIP fractionation at the transition region. We also observe a gradual evolution over time and structures in the AR. 

A broader motivation for this work arises from the fundamental question of wave origin in the context of the ponderomotive force model of FIP fractionation. Given this model, the next key step is to determine where the responsible waves originate. The distinction between resonant and nonresonant waves, as inferred from the behavior of sulfur, appears to favor resonant waves and thus supports a coronal origin. A more direct line of reasoning, however, comes from the observations of \cite{widing_rate_2001}, who found that FIP fractionation begins to develop only after coronal loops have extended into the corona. If the source of the relevant waves were located lower in the atmosphere, such as in the photosphere, one would expect the loops to emerge already exhibiting FIP fractionation. The \cite{widing_rate_2001} study remains limited in scope, however, and warrants further investigation and replication.

We suggest that future SPICE SPROUTS investigations prioritize the Mg/Ne and S V/O IV ratios as key diagnostics, even though we draw attention to the potential strong temperature effects on the FIP bias, especially for the magnesium lines.

Coupling SPICE's observations with high-resolution EUV imaging from Solar Orbiter's EUI \citep{2022EUI}, unfortunately unavailable at the time of these observations, could provide insightful material to discriminate regimes of FIP fractionation, i.e. if the plasma observed fractionates low or high in the chromosphere and its associated mechanisms. The SWA/HIS instrument \citep{OwenHIS}, part of SolO's in-situ suite, could also provide connectivity and further confirmation of solar wind streams coming out of the regions studied.

The current study focuses on a specific subset of the SPROUTS observations to establish a robust methodological framework for these novel observations. Future studies will build upon this foundation and incorporate data from other solar observatories. In future work, we plan to extend our analysis to include a broader range of data sources pertaining to this AR and the slow solar wind possibly emanating from its eastern boundary.

Further investigation into the evolution of this AR can be achieved by utilizing observations from other missions. The Extreme-Ultraviolet Imaging Spectrometer \citep{EISInstrument} on board the Hinode \citep{Hinode} mission observed the AR on the days following the SPICE observations. Measurements of Doppler velocities, densities, and other abundance diagnostics could provide additional context as well as insights into the AR's continued evolution. Using the Polarimetric and Helioseismic Imager (PHI) \citep{PHIinstrument} and advanced modeling, we could extrapolate the coronal magnetic field. This would allow us to better visualize the magnetic structure of the AR itself and understand its interrelationship with the structures observed in the EUV images, as the Sun's magnetic field drives the dynamics and structure of the solar corona. Magnetic field extrapolations also help determine where the plasma reaching Solar Orbiter originates, as shown in Figure B3 in the appendix. Combining these model predictions with the analysis of in-situ data would give us a clearer idea if the plasma from this AR would eventually reach the spacecraft through reconnection at the boundary with the coronal hole to its south. Notably, SWA-HIS \citep{OwenHIS}, which measures heavy ion abundances in the solar wind, recorded data on the days following these spectroscopic observations and the data is now available. 
In addition to these efforts, other data sets of SPROUTS observations will allow us to further explore temporal changes in a wider variety of solar structures.

\begin{acknowledgements}
The authors would like to thank the referee for their very insightful comments and suggestions.
These efforts at SwRI for Solar Orbiter SPICE are supported by NASA under GSFC subcontract \#80GSFC20C0053 to Southwest Research Institute. The development of the SPICE instrument has been funded by ESA member states and ESA (contract no. SOL.S.ASTR.CON. 00070). J.M.L. was funded by basic research funds of the Office of Naval Research. N.Z.P. is supported by STFC Consolidated Grant ST/W001004/1.
Solar Orbiter is a space mission of international collaboration between ESA and NASA, operated by ESA. The development of SPICE has been funded by ESA member states and ESA. It was built and is operated by a multi-national consortium of research institutes supported by their respective funding agencies: STFC RAL (UKSA, hardware lead), IAS (CNES, operations lead), GSFC (NASA), MPS (DLR), PMOD/WRC (Swiss Space Office), SwRI (NASA), UiO (Norwegian Space Agency).
This research used the open source \texttt{sospice}\footnote{\url{https://github.com/solo-spice/sospice}} Python package, Fiasco \citep{Barnes2024}, the CHIANTI database version 11 \citep{Dufresne2024}, Astropy and Sunpy.
\end{acknowledgements}


\begin{appendix}
\section{Spectra and Radiance maps}

\begin{figure*}[htbp!]
    \centering
    \includegraphics[width=\linewidth]{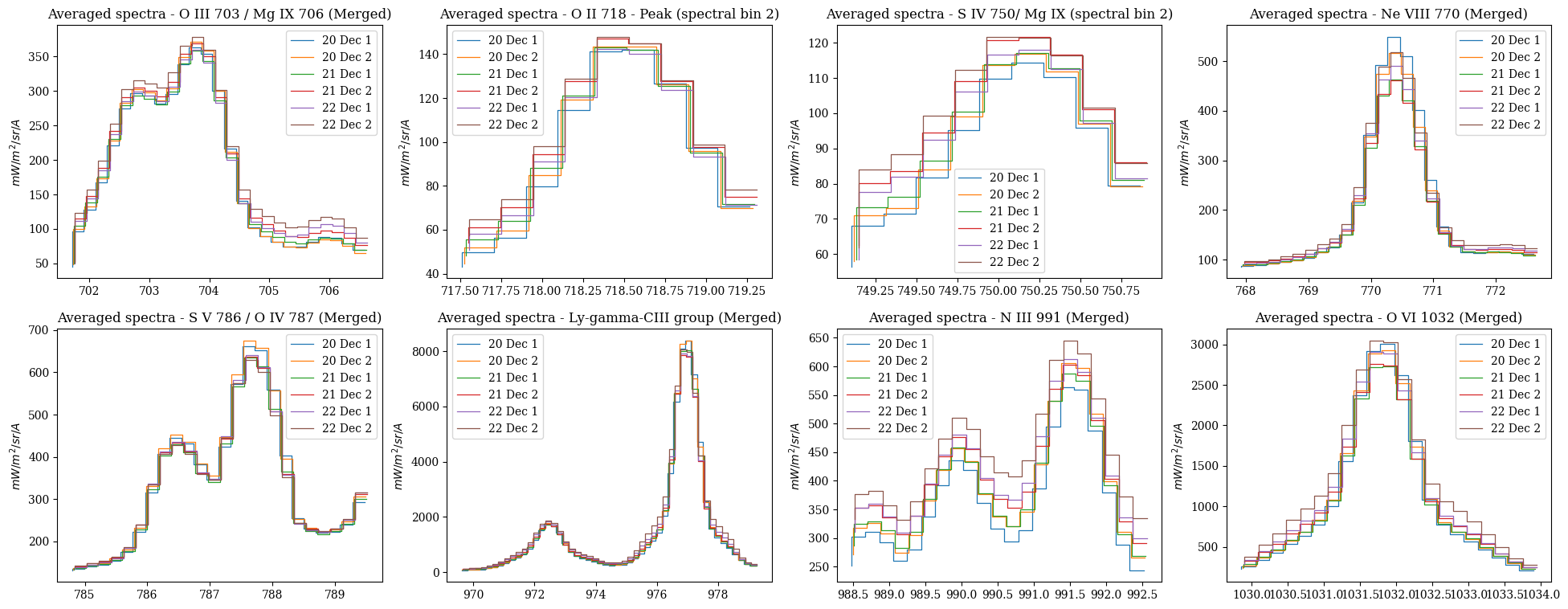}
    \caption{Average spectra over the footpoint of AR 13169, computed over an area of 15x30 pixels. Each color represents a different observation time.}
    \label{fig:avspecDec2022}
\end{figure*}

\begin{figure*}[htbp!]
    \centering
    
    \includegraphics[height=14.5cm]{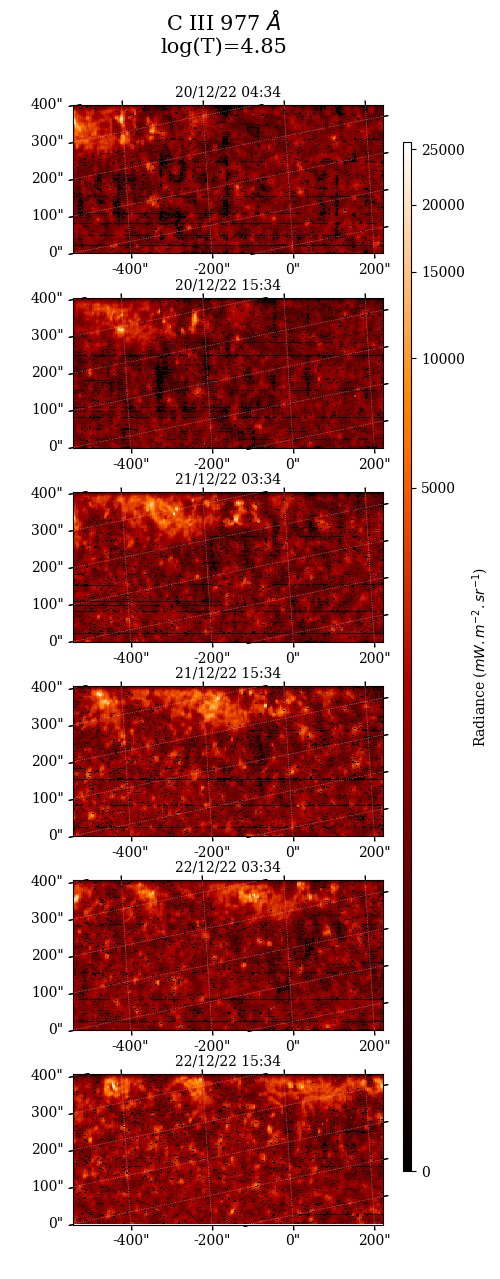}
    \includegraphics[height=14.5cm]{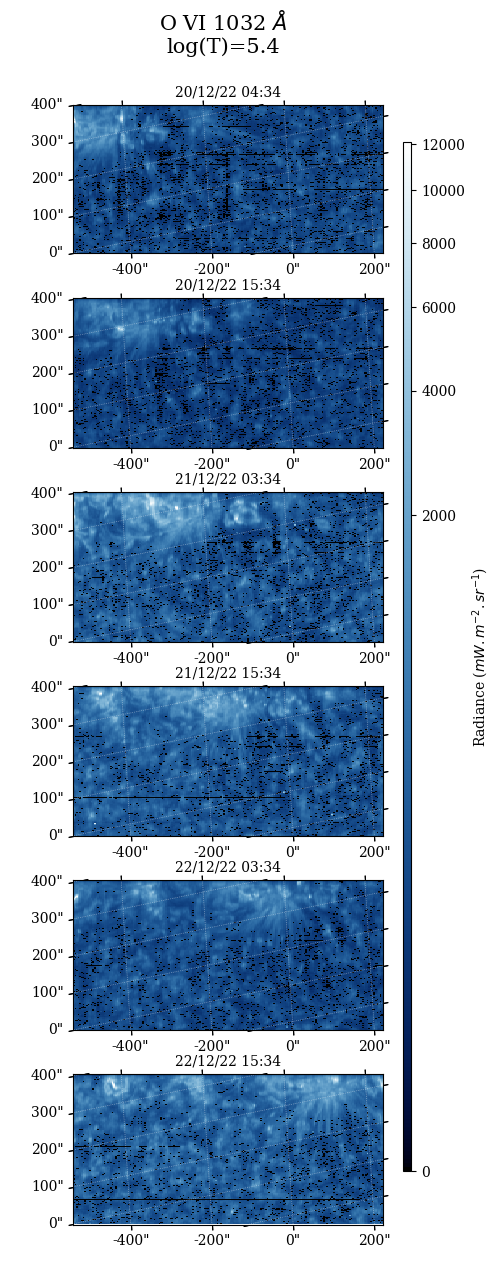}
    \includegraphics[height=14.5cm]{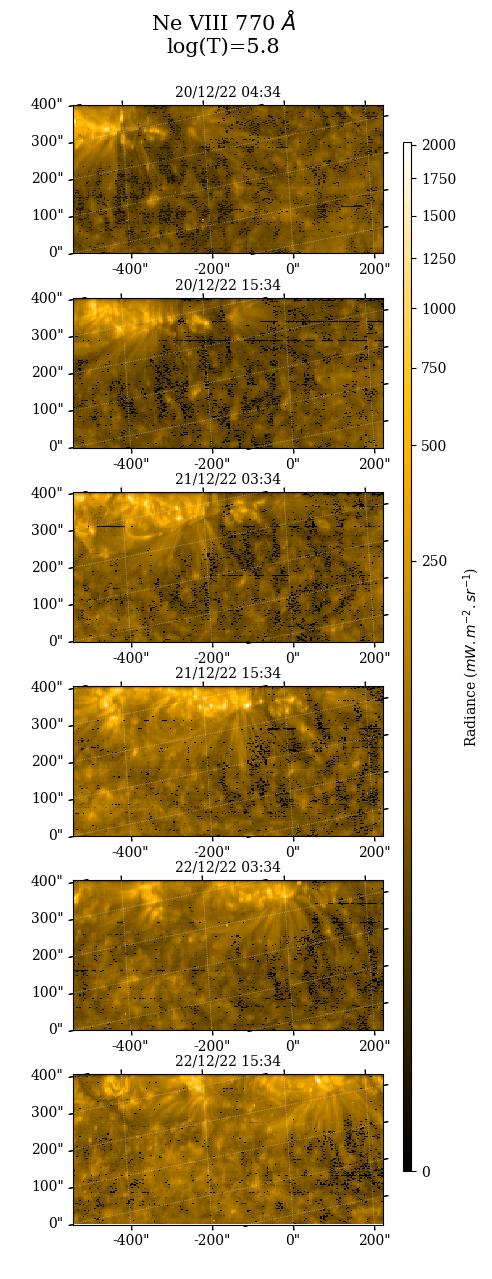}
    
    \caption{Radiance maps closeups seen in the C III 977 \AA, O VI 1032 \AA\;and Ne VIII 770 \AA\;lines from December 20-22 2022. The field of view is focused on AR 13169 (upper parts of the panels). Time progresses as shown in Figure \ref{fig:FIP_bias_time}.}\label{fig:radmapDec2022}
    
\end{figure*}

\section{Uncertainty computation} \label{sec:uncertainties}

\subsection{Radiance uncertainty}

We focus here on the lines that were used for the FIP bias computation, but the method applies for every lines. For the \ion{O}{iii} 703 \AA\;/ \ion{Mg}{ix} 706 \AA,  \ion{S}{v} 786 \AA\;/ \ion{O}{iv} 787 \AA\;and \ion{N}{iii} 991 \AA\;window, we used triple Gaussian fitting with a uniform background. All centroids, amplitudes, and widths of the lines were free parameters with bounds, as well as the uniform background, since this set of three blended lines typically has enough signal-to-noise ratio for the lines to be easily identified and separated. The \ion{Ne}{viii} 770 \AA\;/ \ion{Mg}{viii} 772 \AA\;and \ion{C}{iii} 977 \AA\;window have been fitted with two gaussians, on the same methodology as the three-gaussian window.


We compute the radiances by passing initial guesses to a bounded nonlinear least squares fitting routine (\texttt{curve\_fit}). From the best-fit parameters, the integrated radiance is calculated using the formula for the area under a Gaussian curve:
\[
\mathrm{Radiance} = \frac{A}{2} \sqrt{2\pi} \cdot \sigma
\]
where $A$ is the fitted amplitude and $\sigma$ the standard deviation. This approach ensures a consistent and robust estimate of the total emission in each spectral feature, which is critical for the subsequent derivation of temperature diagnostics and FIP bias.

The uncertainties associated with radiances are estimated by propagating the errors from the spectral line fitting through each step of the calculation. Instrument-related errors from the \texttt{sospice}\footnote{\url{https://github.com/solo-spice/sospice}} package are embedded in the line fitting routine with the \texttt{sigma} parameter of the \texttt{curve\_fit} routine. The radiance uncertainty for each fitted spectral line is then derived from the covariance matrix output. Specifically, the uncertainty on the integrated line intensity is computed using the formula for error propagation of a Gaussian:
    \[
    \delta I = \sqrt{2\pi} \left[ (\sigma \cdot \delta A) + (A \cdot \delta \sigma) \right]
    \]
    where $A$ is the line amplitude, $\sigma$ is the Gaussian width, and $\mathrm{cov}_{A,\sigma}$ is their covariance. 

\subsection{DEM uncertainty}

The DEM forward problem is the same as in \cite{Plowman_2020}. $D_i$ are the data and $\sigma_i$ their associated uncertainties. $R_i$ are the contribution functions. $M_i$ is the model fit to the data, and $c_j$ are the unknown coefficients of the DEM solution. Here we discuss the DEM uncertainty estimation method used in this paper which was first made public in the EMToolKit Github \citep{Plowman_EMToolKit_A_Standardized}.

To estimate the level of uncertainty in the DEM estimation, we need to evaluate the following: given a solution that satisfies $M_i=D_i$, how much can the coefficients $c_j$ vary -- resulting in a perturbed model $M'_i = \sum_j R_{ij}(c_j + \Delta c_j)$ -- before the modified solution reaches a given threshold $\chi'^2$, which is typically the number of data points (`reduced' $\chi'^2=1$).

This question can have multiple interpretations, each yielding a different estimate of uncertainty. One may ask how much a single coefficient $c_j$ can change, or how much all coefficients can vary collectively. The parameters could change coherently (e.g., all increasing or decreasing together), incoherently (e.g., in random directions), or in ways that compensate for each other. These scenarios produce very different results: fully compensating changes may yield formally infinite or undefined uncertainties, while coherent variation produces relatively small uncertainties. Incoherent or random variation lies in between and generally scales with the square root of the number of parameters.

For a more conservative estimation, a "one-parameter-at-a-time" approach is adopted: each $c_j$ is varied individually while holding others fixed. This yields the most straightforward and robust uncertainty estimate without requiring detailed knowledge of the solution. 

Let $\Delta c_j$ be the only source of deviation from agreement (meaning that the other $c$ in the coefficient vector stay constant) then $\chi'^2=n_\mathrm{data}$ implies that $$\sum_i \frac{(R_{ij}\Delta c_j)^2}{\sigma_i^2} = n_\mathrm{data}$$
We can then estimate the errors as $$\Delta c_j = \frac{\sqrt{n_\mathrm{data}}}{\sqrt{\sum_i R_{ij}^2/\sigma_i^2}}$$

Considering original contribution functions, $R_{ij} = \int R_i(T) B_j(T)\,dT $ 
For the uncertainty estimate, we take the basis functions to be top hats of width $\Delta \log_{10} T=0.1$ K at temperature $T_j$ (this is distinct from the width of the basis functions used for estimating the DEM, and should generally be taken to be of order the temperature resolution of the instrument). Thus $R_{ij} = R_i(T_j)\Delta \log_{10} T$ and the uncertainty in a DEM component with that characteristic width is:
$$\sigma_{E_j} = \frac{\sqrt{n_\mathrm{data}}}{\sqrt{\sum_i (R_i(T_j) \Delta \log_{10} T)^2/\sigma_i^2}}$$

$\sigma_{E_j}$ is the uncertainty estimate we use in this paper. Correlated fluctuations in unresolved temperature elements can of course be higher, as is also the case in (for example) imaging observations.

\subsection{FIP bias uncertainty}
The FIP bias uncertainty is estimated by propagating the uncertainties in the intensities and contribution functions through the FIP bias equation (see Eq. \ref{eq:fip}). 
Since the FIP bias is computed on the average spectrum over a given zone, the radiance and DEM uncertainties are normalized by the size of the zone.
The propagated error $\delta\mathrm{FIP}$ on the relative FIP bias is computed as:
\[
\delta\mathrm{FIP} = \mathrm{FIP}\cdot \sqrt{\left(\frac{\delta I_{LF}}{I_{LF}}\right)^2+\left(\frac{\delta I_{HF}}{I_{HF}}\right)^2+\left(\frac{\delta D_1}{D_1}\right)^2+\left(\frac{\delta D_2}{D_2}\right)^2}
\]
With:
\begin{equation} 
\begin{split}
D_1 & = \; <C_{HF} \cdot \text{DEM}> \\
D_2 & = \; <C_{LF} \cdot \text{DEM}> \\
\delta D_1 & =  \sqrt{\sum_i\left(C_{HF_i}^2 \cdot \delta \text{DEM}_i^2 + \delta C_{HF_i}^2 \cdot \text{DEM}_i^2 \right) } \\
\delta D_2 & =  \sqrt{\sum_i\left(C_{LF_i}^2 \cdot \delta \text{DEM}_i^2 + \delta C_{LF_i}^2 \cdot \text{DEM}_i^2 \right) } \\
\end{split}
\end{equation}
The error margin on the contribution functions ($\delta C_{HF/LF_i}$) has been taken as 15\% of their value.

This approach incorporates the uncertainties in radiance measurements, DEMs and contribution functions into the final FIP bias uncertainty.

\end{appendix}

\bibliography{bib}{}
\bibliographystyle{aasjournal}

\end{document}